\documentclass[11pt,a4paper]{article}
\usepackage{jheppub,amsmath,  amssymb,slashed,url,bm,textgreek,upgreek}
\usepackage{graphicx}
\usepackage{epstopdf}
\def\t{{ \sf t}} 
\def\tt{{\mathfrak t}}

\def\be{\begin{equation}}
\def\ee{\end{equation}}

\def\eff{{\mathrm{eff}}}

\def\h{\widehat}

\def\y{{\mathrm y}}

\def\D{{\mathcal D}}

\def\O{{\mathcal O}}

\def\d{{\mathrm d}}

\def\R{{\mathbb R}}

\def\D{{\mathcal D}}
\def\[{\bigl [}

\def\]{\bigr ]}

\def\Z{{\mathbb Z}}

\def\t{\widetilde }
\def\h{\widehat}

\def\la{\langle}
\def\ra{\rangle}

\font\teneurm=eurm10 \font\seveneurm=eurm7  \font\fiveeurm=eurm5
\newfam\eurmfam
\textfont\eurmfam=\teneurm \scriptfont\eurmfam=\seveneurm
\scriptscriptfont\eurmfam=\fiveeurm

\font\teneusm=eusm10 \font\seveneusm=eusm7 \font\fiveeusm=eusm5
\newfam\eusmfam
\textfont\eusmfam=\teneusm \scriptfont\eusmfam=\seveneusm
\scriptscriptfont\eusmfam=\fiveeusm

\font\tencmmib=cmmib10 \skewchar\tencmmib='177
\font\sevencmmib=cmmib7 \skewchar\sevencmmib='177
\font\fivecmmib=cmmib5 \skewchar\fivecmmib='177
\newfam\cmmibfam
\textfont\cmmibfam=\tencmmib \scriptfont\cmmibfam=\sevencmmib
\scriptscriptfont\cmmibfam=\fivecmmib

\def\Tr{{\mathrm{Tr}}}
\def\i{{\mathrm i}}
\def\d{{\mathrm d}}
\def\eff{{\mathrm{eff}}}
\def\x{{\sf x}}
\def\y{{\sf y}}
\title{Duality and Axion Wormholes}

 \author{Edward Witten}
\affiliation{School of Natural Sciences, Institute for Advanced Study,\\ 1 Einstein Drive, Princeton, NJ 08540 USA}
\abstract{The prototype of a Euclidean wormhole solution of Einstein gravity coupled to matter is the axion wormhole in four spacetime dimensions.  In this primarily expository article,
we spell out some details about this construction.    The axion wormhole has a semiclassical description,
found in the original paper \cite{GS}, in which the matter system is a two-form gauge field $B$ with three-form field strength $H=\d B$.   The two-form is dual to a massless scalar,
but the wormhole does not have a semiclassical description in terms of the scalar.   There is no contradiction here as the duality between the two-form and the scalar is not a simple transformation of
classical fields but involves, in Euclidean signature, a Poisson resummation of the sum over fluxes.   Because of the need for  this Poisson resummation, the scalar field cannot be treated
semiclassically in the wormhole throat.   Nonetheless, it is straightforward to compute the effective action derived from the wormhole in the scalar (or two-form) language, recovering
standard claims.
}

\begin{document}\maketitle

\section{Introduction}\label{intro}

\begin{figure}
 \begin{center}
   \includegraphics[width=3.5in]{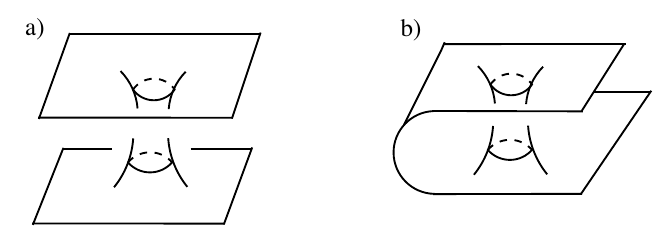}
 \end{center}
\caption{\footnotesize  (a) A wormhole connecting two asymptotically flat worlds.  (b)  A wormhole that provides a shortcut between distant regions of the same
asymptotically flat world.}\label{Tube}
\end{figure}

A wormhole is, for example,  a geometrical connection between two different asymptotically flat worlds, or a shortcut between otherwise distant regions of the same
asymptotically flat world  (fig. \ref{Tube}).    It is natural to look for wormholes as classical Euclidean  solutions of Einstein gravity coupled to a suitable matter system,
but in practice very few examples are known.   The prototypical example is the ``axion wormhole'' 
  \cite{GS} in which the matter system is  a two-form field $B=\frac{1}{2}\sum_{i,j} B_{ij} \d x^i \d x^j$, with field
strength $H=\d B=\frac{1}{3!}\sum_{i<j<k}\d x^i \d x^j \d x^k H_{ijk}$, $H_{ijk}=\partial_i B_{jk}+\partial_j B_{ki}+\partial_k B_{ij}$, and coupling parameter $h$.
The action in Euclidean signature is
\be\label{action}I=\int \d^4x \sqrt g\left(-\frac{R}{16\pi G}+\frac{1}{2\cdot 3!\, h^2}H^{ijk} H_{ijk} \right)  .\ee   This system has a simple classical wormhole solution with the topology
suggested in fig. \ref{Tube}(a); for sufficiently small $h$, the curvatures and field strengths in the solution are small and the solution is semiclassically reliable.
There are significant generalizations with the addition of scalar fields and possibly with additional two-forms   (for example, see \cite{AOP,Berg}), but the basic example will suffice for our purposes.

The reason that the wormhole solution of the system (\ref{action}) is known as an ``axion'' wormhole is that in fact, in four dimensions, the two-form field $B$ is dual to a scalar
field $\phi$, the ``axion.''   As detailed in section \ref{duality}, there  is a complete equivalence of two theories
\be\label{dualityz} \frac{1}{2 f^2}\int\d^4x\sqrt g\,\partial_i \phi\partial^i\phi \leftrightarrow  \frac{1}{2\cdot 3! h^2}\int \d^4x \sqrt g H^{ijk} H_{ijk}  ,\ee
provided an appropriate relation is imposed between the constants $f$ and $h$ and provided $\phi$ and $B$ both satisfy Dirac quantization.   Here Dirac
quantization means that periods of the one-form
$\d\phi$ and the three-form $H$ are required to be integer multiples of $2\pi$.

Since this duality holds in any background metric, one expects, after coupling to gravity, an equivalence of two theories:
\be\label{twoth}\int\d^4x\sqrt g\left(-\frac{R}{16\pi G} +  \frac{1}{2 f^2}\partial_i \phi\partial^i\phi\right)\leftrightarrow   \int \d^4\sqrt g\left(-\frac{R}{16\pi G}+\frac{1}{2\cdot 3! h^2}H^{ijk} H_{ijk} \right). \ee
Despite this equivalence, the two-form theory has a real classical wormhole solution and the scalar theory does not.   This was observed in the original paper \cite{GS} and can be understood by
a simple scaling argument.   If one scales by metric by $g\to e^s g$, with a  real constant $s$, then in the two-form theory, the two terms $\sqrt g R$ and $\sqrt g H^2$ in the action
scale oppositely,
so a balance between them is possible, leading to a classical solution with a ``throat'' of a definite size.
But in the scalar theory, the two terms $\sqrt g R$ and $\sqrt g \partial_i\phi \partial^i\phi$ scale in the same way, so a balance  
is not possible, suggesting that a classical solution does not exist and showing that if a solution does exist, its action will vanish and the size of the throat will  be arbitrary.

The equivalence of the two theories is not in conflict with  the fact that the wormhole exists as a classical solution in one description but not in the other.
Indeed, in Euclidean signature the duality between the scalar and the two-form is far from being a simple map between classical solutions.\footnote{As explained in section
\ref{duality}, except for important subtleties involving zero-modes,
in Lorentz signature the duality can be understood as a map between classical solutions.}    A Poisson resummation is involved,
as essentially first discussed in this context in \cite{KLC} and further considered, for example, in  \cite{AOP}.    The main purpose of the present article is to spell out some consequences of this
fact.   These consequences, though not fundamentally new, have been somewhat implicit in the existing literature.     

The main point we aim to elucidate is as follows.
For sufficiently small $h$, as observed in the original paper \cite{GS}, there is a reliable semi-classical wormhole solution in the two-form description.   Since the two-form theory is dual to the scalar
theory, the same wormhole also exists in the scalar theory.   But because of the Poisson resummation that is involved in relating the two-form to the scalar, the scalar field cannot be treated
semiclassically in the wormhole throat.  We want to understand how to study the wormhole in the scalar language and in particular how to compute the effective action in that language.
 
 The fact that the wormhole has a semiclassical description for small $h$ in the two-form language suggests the following question: is there a parameter
region, perhaps small $f$, in which the wormhole has a semiclassical description in the scalar language?   The answer to this question is ``no,'' 
since the scalar version of the theory does not have a physically sensible wormhole solution, semiclassically reliable or not,  as suggested by a scaling argument that was mentioned earlier  and
 as shown in the original paper \cite{GS}.
 
The statement that the scalar theory does not have a physically sensible wormhole solution requires  some elaboration.  Indeed, some confusion surrounds this point.
In Euclidean signature, the  classical equations of motion of the two-form theory can actually be mapped to those of the scalar theory by setting
\be\label{zifflo} \partial_k \phi=-\i\frac{f}{3!h}\epsilon_{klmn} H^{lmn},~~~\i=\sqrt{-1}.  \ee    (The precise factor $-\i f/3!h$ ensures that  the stress tensor transforms properly, leading to  an equivalence
of the Einstein field equations coupled to $\phi$ or $B$.  The overall sign is an arbitrary convention.)
In Lorentz signature, a similar relation without the factor of $-\i$ does largely, but not fully, account for the duality, from a Hamiltonian  point of view, 
as explained in section \ref{duality}.    This is not the case for path integrals.
The relation (\ref{zifflo}) implies the classical equations of motion for both $\phi$ and $H$, and hence cannot be defined off-shell.  So  it does not make sense as a transformation
of the fields in a path integral, which are generically off-shell. 
But  eqn. (\ref{zifflo}) can be deduced from path integrals as a statement that holds inside  quantum correlation functions.  This will be explained  in section \ref{duality}.

Even on-shell, there is a problem with interpreting eqn. (\ref{zifflo}) as a relation between classical fields.
Dirac quantization requires $\d\phi$ and $H$ to have real periods (indeed, periods that are integer multiples of $2\pi$), but the factor of $-\i$ in  (\ref{zifflo}) makes
this impossible if (\ref{zifflo}) is regarded as a relation between classical fields. Even in Lorentz signature, where there is no factor off $-\i$ in the formula, the factor of
$f/h$ means that as a relation between classical fields, eqn. (\ref{zifflo}) is generically incompatible with Dirac quantization.

In the particular case of the wormhole, since the three-form $H$ has a non-zero (and real!) flux when integrated on a cross-section of the wormhole throat, the relation
(\ref{zifflo}) would require $\phi$ to change by an imaginary constant in going ``through the wormhole'' from one asymptotically flat region to the other.   In general, in the gravitational path
integral, one considers classical fields that are not real, but one does want to impose suitable reality conditions in asymptotic regions of spacetime.  In the case of the $\phi$ field,
one wants $\phi$ to equal a real constant at infinity; the value of this constant represents a choice of vacuum.   If $\phi$ changes by an imaginary amount in going through the wormhole,
it cannot be real at both ends, and hence the solution obtained by applying the transformation (\ref{zifflo}) to the wormhole solution of \cite{GS} does not satisfy the boundary conditions that
one would  want in wormhole physics.

 Indeed, it would be highly problematical  to give $\phi$ an imaginary value at infinity.  To see this, consider elaborating the theory slightly by adding gauge
fields with field strength $F$ and an axion coupling that in Euclidean signature is 
$\i\int \phi \,\Tr\, F\wedge F$ (in Lorentz signature there is no factor of
$\i$ in this coupling).   If $\phi$ has an imaginary expectation value, the gauge fields have an imaginary theta-angle, violating unitarity.

Despite all this, in more complicated theories with one or more two-forms coupled to scalars, the relationship (\ref{zifflo}) and its generalizations are very useful; they have been used to
efficiently analyze wormhole solutions of low energy effective actions derived from string theory  \cite{AOP}.   In general, as shown in that paper,
a very efficient way to find wormhole solutions of a theory with multiple scalars and two-forms is to convert all of the two-forms to imaginary scalars by relations such as (\ref{zifflo}),
solve the equations of motion in the scalar language, and then convert the imaginary scalars back to two-forms via (\ref{zifflo}).

The contents of this paper are as follows.
In section \ref{duality}, we analyze the duality between a scalar and a two-form in four dimensions (some further details about this are explained in 
two appendices).  In section \ref{wormhole}, we describe the wormhole solution of \cite{GS}.   In section \ref{effac}, we describe the effective action
associated to the wormhole.      Indeed, the most salient property of the wormhole in quantum theory is that it appears to generate a bilocal effective action,
with the bilocality coming from an integration over the positions of the two ends of the wormhole \cite{KS}; this is related to the claim that the wormhole generates
some randomness in ``constants'' of nature \cite{Coleman}.    The bilocal effective action of the wormhole
is relatively obvious in the two-form language, and we show how it can also be derived in the scalar language, even though the scalar field cannot be treated
semiclassically in the wormhole throat.  

This article is largely expository and most of the content is not really novel.
The literature on dualities in theories with abelian gauge and/or global symmetries is too vast to properly cite here.
The derivation in section \ref{derivation} of the duality between the scalar and the two-form is in the spirit of various previous derivations \cite{Bu,VR,Witten}.
The duality between a scalar and a two-form in four dimensions has been previously studied in \cite{ST,Barbon} and in considerable detail in \cite{DMW}. 
Additional useful 
references on mathematical background include \cite{FMS,MooreSaxena}.

\section{The Duality}\label{duality}

\subsection{Dirac Quantization and Relation Between Couplings}\label{dirac}

We consider a two-form and a scalar with the actions defined in eqn. (\ref{dualityz}).
Before giving a formal proof of the duality between these fields, we will give a qualitative explanation, starting in Lorentz signature.

The relation\footnote{In Euclidean signature, the following relation
requires an extra factor of $\pm\i$; see eqn. (\ref{zifflo}).}
\be\label{noflo}\partial_i \phi=\frac{f}{3!h}\epsilon_{ijkl} H^{jkl}  \ee
almost defines an on-shell transformation between a $B$-field with field strength $H=\d B$ and a massless scalar $\phi$.
This is most obvious if the transformation is written in terms of differential forms.  In that language, eqn. (\ref{noflo}) reads $\d \phi=\frac{f}{h}\star H$, where $\star $ is
the Hodge star operator.   It exchanges two equations satisfied by $\phi$, namely the classical equation of motion $\d\star\d\phi=0$ and the identity $\d^2\phi=0$,
with two equations satisfied by $H$, namely the Bianchi identity $\d H=0$ and the classical equation of motion $\d\star H=0$. Indeed, if $\d\phi=\frac{f}{h}\star H$,
then the identity $\d^2\phi=0$ is equivalent to the classical equation $\d\star H=0$, and the classical equation $\d\star\d \phi=0$ is equivalent to the Bianchi identity $\d H=0$.
The constant $f/3!h$ in eqn. (\ref{noflo}) is chosen so that if $\phi$ and $H$ are related  as in eqn. (\ref{noflo}), then the stress tensor of $\phi$ equals the stress tensor of $H$.

Because it exchanges the equations of motion with the Bianchi identities $\d(\d\phi)=\d H=0$, the transformation (\ref{noflo}) implies the equations of motion and cannot
be defined  off-shell.  So it does not make sense as a transformation of the classical fields in a path integral, which are generically off-shell.
(As we will see, it can be deduced from path integrals as a statement about quantum correlation functions; roughly, from the point of view of path integrals,
 this relation holds on the average.)  In a Hamiltonian approach, one constructs a Hilbert space by quantizing
the classical phase space of a theory.  The classical phase space can be viewed as the space of classical solutions of a theory, so in a Hamiltonian approach, one never
has to consider off-shell fields; one only has to consider on-shell fields and their canonical commutation relations.   
The relation (\ref{noflo}) does map the canonical commutation relations of $\phi$  to those of $B$.    So, in a Hamiltonian approach,
(\ref{noflo}) almost suffices to characterize a quantum duality between
the scalar and the two-form.   But eqn. (\ref{noflo}) does not define a complete quantum duality in a Hamiltonian approach,  because it omits 
important information about zero-modes on the two sides.   If $H$ is given, then $\phi$ is only determined up to an additive constant.   And if $\phi$ is given, then eqn. (\ref{noflo}) determines
$H$, but this does not quite suffice to determine $B$ up to a gauge transformation, because there can be $B$-fields that are not pure gauge, but have $H=0$ (for example, a $B$-field on
a two-manifold $\Sigma_2$  with $\int_{\Sigma_2}B\not=0$).   
To extend the duality to encompass the zero-modes, one has to impose Dirac quantization on both sides of the duality, and impose a certain relation between $f$ and $h$.
   Moreover, for the zero-modes, the duality cannot be expressed by
a classical mapping like (\ref{noflo}) between scalars and two-forms, but involves quantization.

For $\phi$, Dirac quantization means that $\phi$ is not a real-valued field, but is valued in a circle, which we will take to have circumference $2\pi$.   Accordingly, the one-form $\d\phi$
can have nontrivial periods, but these periods are required to be integer multiples of $2\pi$.   Thus, for an oriented circle $S$ in spacetime,
\be\label{zimbo}\oint_S \d\phi\in 2\pi\Z. \ee
For $B$, Dirac quantization amounts to two conditions, which are in parallel with the conditions satisfied by $\phi$.
First, if  $\Sigma_2$ is an oriented two-cycle in spacetime, then the ``holonomy'' $\int_{\Sigma_2} B$ is only gauge-invariant mod $2\pi$
(analogously to the fact that $\phi$ is only well-defined mod $2\pi$).  This may be stated as follows:   when we make a gauge transformation by $B\to B+\d\Lambda$,
$\Lambda$ must be understood as a $U(1)$ gauge field (whose curvature $\d\Lambda$ 
 may have nontrivial periods that are integer multiples of $2\pi$), not simply as a one-form, so this gauge transformation can shift $\int_{\Sigma_2}B$ by an integer 
multiple of $2\pi$.
 Second, the periods of $H=\d B$ are integer multiples of $2\pi$, so that if $\Sigma_3$ is an oriented three-cycle in
spacetime, then
\be\label{wimbo}\int_{\Sigma_3}H\in 2\pi\Z. \ee 
These conditions mean that $B$ is a two-form analog of a gauge field with gauge group $U(1)$, rather than $\R$.

Simple examples will suffice to show that Dirac quantization is needed in order to make the duality work.   Let us formulate our theory 
on a spacetime $M=\R\times W$, where the first factor is parametrized by the ``time'' $t$ and the second is a compact three-manifold.
On $M$ we pick a 
time-independent metric
$\d s^2=-\d t^2+ \sum_{i,j=1}^3 h_{ij}\d x^i \d x^j$, where $h_{ij}$ is a metric on $W$, say of volume $V$.  Let $\varphi(t)$ be the ``constant'' mode of $\phi$ that depends only on $t$.
  The action for this mode is
$ I_{\varphi}=\frac{V}{2f^2}\int \d t \dot\varphi^2$, with $\dot\varphi=\frac{\d\varphi}{\d t}$.   Introducing the canonical momentum $p_\varphi=\delta I_\varphi/\delta\dot\varphi= V\dot\varphi/f^2$, the corresponding Hamiltonian is
\be\label{hammode} H_\varphi=\frac{f^2}{2V}p_\varphi^2. \ee
If $\phi$ is a real-valued field, then $\varphi$ is also real-valued.   Then $H_\varphi$ is the Hamiltonian of a free particle in one dimension, and has a continuous spectrum.   
This continuous spectrum will not match the quantization of the $B$-field in the same geometry.   Suppose, however, that $\phi$ and $B$ satisfy Dirac quantization.   Then
$\varphi$ is circle-valued, $\varphi\cong \varphi+2\pi$, so $p_\varphi$ has integer eigenvalues $q\in \Z$, and the eigenvalues of $H_\varphi$ are of the form 
\be\label{mellow}E_q=\frac{f^2 q^2}{2V}.\ee
To match this on the dual side, we note that the two-form theory has classical solutions in which $H$ is a  time-independent  multiple of the Levi-Civita volume
form $\Omega_W$ of the three-manifold $W$.
Dirac quantization says that we must have $\int_W H=2\pi q$ for some integer $q$, and therefore $H$ must be precisely $2\pi q\Omega_W/V$.   The energy of this field is 
\be\label{pillow} E'_q=\frac{(2\pi q)^2}{2Vh^2}. \ee
We see that $E_q=E'_q$ if and only if 
\be\label{zillow}h=\frac{2\pi}{f}, \ee
and therefore the duality requires this relation as well as  Dirac quantization of $\phi$ and $B$.

For another example, we can consider a winding mode of $\phi$.   Specialize to the case that $W=S\times N$, where $S$ is a circle of circumference $L$, and $N$ is a two-manifold of volume
$V'$.
We assume on  $M$ a product metric  $-\d t^2+\d \psi^2+ \sum_{a,b=1}^2k_{ab} \d y^a \d y^b$, where $\psi\cong \psi+L$ parametrizes $S$ and the last term is a metric
on $N$.   We can now consider a classical solution for $\phi$ with $\phi=2\pi \psi r/L$, with an integer $r$.  Thus as $\psi$ varies over the interval $[0,L]$,
 $\phi$ winds  around $S$ with winding number $r$.
 The energy of this winding solution is
\be\label{illow}\t E_r=\frac{(2\pi r)^2V'}{2f^2 L}. \ee
To match this on the dual side, we consider a mode of the $B$-field of the form $\beta(t)\Omega_{N}/V'$, where $\Omega_N$ is
 the Levi-Civita form of $N$; the normalization was chosen so that the period of $B$ is $\int_N B= \beta(t)$.   
The action for this mode is $\frac{L}{2h^2 V'} \int \d t \,\dot\beta^2$, so the canonical momentum is $p_\beta=\frac{L\dot\beta}{h^2 V'}$ and the Hamiltonian is 
$H_{\beta}=\frac{V'h^2 p_\beta^2}{2 L}$. 
 As explained earlier, Dirac quantization means that $\beta$ is an angular variable with period $2\pi$, so the canonical
momentum $p_\beta$ has integer eigenvalues.  A state with $p_\beta=r$ has energy
\be\label{nillok} \t E'_r= \frac{V' h^2r^2}{2L}.  \ee
We see that to get $\t E_r=\t E'_r$, we again need $h=2\pi/f$.    

A systematic discussion of these matters (in the greater generality of a $p$-form gauge field and an $n-p-1$-form gauge field in $n$ dimensions) can be found in 
\cite{MooreSaxena}.

\subsection{Derivation of the Duality}\label{derivation}

We will now give a formal derivation of the duality between the scalar and the two-form.   The first step is to observe that the circle-valued massless scalar field $\phi$
has a $U(1)$ symmetry that acts by shifting $\phi$ by a constant.   We introduce a $U(1)$ gauge field $A$ that gauges this symmetry;
the gauge invariance will be $\delta\phi=\varepsilon,$ $\delta A_i=-\partial_i\varepsilon$. 
The field strength  of $A$ is as usual
$F_{ij}=\partial_i A_j-\partial_j A_i$; it  can be packaged in the two-form  $F=\d A=\frac{1}{2}\sum_{i,j} \d x^i \d x^j F_{ij}$.
As a $U(1)$ gauge field, $A$ satisfies Dirac quantization.   
  The gauge-invariant action is\footnote{Geometrically, the circle-valued field $\phi$ is a trivialization of the complex line bundle whose connection is $A$; if $\phi$
  is globally defined, then $A$ must be topologically trivial.   Our point of view in the following is that $A$ is {\it a priori} an arbitrary $U(1)$ connection, not necessarily
  topologically trivial, but the path integral over $\phi$ vanishes if $A$ is not topologically trivial, since the space of $\phi$-fields over which one would want to integrate is empty.
  This viewpoint seems to make the following discussion as elegant as possible and provides a starting point for generalizations in which $\phi$ is replaced by
  a matter system that does not necessarily trivialize $A$.}
\be\label{welgo}\frac{1}{2f^2}\int\d^4x \sqrt g (\partial_i\phi+A_i)(\partial^i\phi+A^i). \ee
Of course, this theory as it stands is not equivalent to the original theory without $A$.    However, we can construct a theory that is equivalent to the original theory by adding another
field $B$ with the property that the path integral over $B$ will set $A=0$ up to a gauge transformation.   There is a standard way to do this: $B$ must be a two-form gauge field
obeying Dirac quantization; thus, as explained in section \ref{dirac}, $B$ will admit the gauge invariance $B\to B+\d \Lambda$, where $\Lambda$ is a $U(1)$ gauge field, and $H=\d B$ has
periods that are integer multiples of $2\pi$.   The coupling between $A$ and $B$ is taken to be (in Euclidean signature)
\be\label{zingo} \frac{\i}{2\pi} \int_M A\wedge H =\frac{\i }{2\pi}\int_M F\wedge B =\frac{\i }{3!(2\pi)}\int \d^4x\epsilon^{ijkl} A_i H_{jkl},\ee

The first two forms of the action  in eqn. (\ref{zingo}) are related by integration by parts, and the last formula is just a more explicit version of the first one.  In the following, 
we use whichever version of the action is convenient.   
Actually, the first form of the action is valid if $A$ is topologically trivial (and can be viewed simply as a one-form), and the second form is valid if $B$ is topologically trivial (and can be viewed
simply as a two-form).   A completely general version of the
action, which reduces to one of the forms given in eqn. (\ref{zingo}) if $A$ or $B$ is topologically trivial, is given by the theory of  Cheeger-Simons differential characters.  For
a physics-focused introduction, 
see \cite{MooreSaxena} or \cite{HS}.   We also explain in Appendix \ref{direct} a  general definition of this action using cobordism and surgery rather than differential characters.

The claim is now that a path integral over $B$
\be\label{ingo}\int DB \, \exp\left(- \frac{\i }{2\pi}\int_M F\wedge B\right) \ee
gives a delta function that sets $A=0$ up to a gauge transformation.\footnote{More precisely, the integral over $B$ gives a constant multiple of such a delta function. The multiple is 1 if the measure for $B$ is chosen correctly, for example by including a factor $1/2\pi$ in eqn. (\ref{onky}).}
It is obvious that the path integral over $B$ vanishes unless $F=0$, but  
this is not quite enough to set $A=0$ up to a gauge transformation.   For this to work depends on the fact that both $A$ and $B$ satisfy Dirac quantization.
To illustrate the idea, suppose that $M=S^1\times S^3$,  and let $A$ be a flat connection on $M$ with a possibly non-zero value of
$\alpha=\int_{S^1} A$.    As $A$ is a $U(1)$ gauge field,  $\alpha $ is gauge-invariant mod $2\pi$.   Dirac quantization means that  $\int_{S^3} H=2\pi r$ for some integer $r$.
The value of the action is then $\frac{\i}{2\pi}\int_M A\wedge H=\i r \alpha.$
The path integral over $B$ gives
\be\label{monky}\sum_r e^{-\i r \alpha}=\delta(\alpha/2\pi).  \ee
For another illustration of the subtlety involved in the assertion that the path integral over $B$ sets $A=0$ up to a gauge transformation, suppose that $M=S^2\times \t S^2$ is
a product of two-spheres.   Suppose further that $A$ is a pullback from the first factor, with $\int_{S^2}F = 2\pi r$, and that $B$ is a pullback from the second factor, with
$\int_{\t S^2}B=\beta$.    The action is then $\i r\beta$.   Dirac quantization says that $r$ is an integer and $\beta$ should be integrated over the interval $[0,2\pi]$.   The
path integral then gives
\be\label{onky}\frac{1}{2\pi}\int_0^{2\pi} \d\beta e^{-\i r\beta}=\delta_{r,0}. \ee 

In general, if $A$ and $B$ both satisfy Dirac quantization, the topological subtleties always work out nicely, as in the last paragraph, so that the path integral over $B$ gives a delta
function setting $A=0$ up to a gauge transformation.   This assertion is actually equivalent to  Poincar\'{e} duality for Cheeger-Simons differential characters.
See \cite{MooreSaxena, HS}.    See also Appendix \ref{direct}.

So an extended theory of the fields $\phi, A, $ and $B$ with the action
\begin{align}\label{helpme}I(\phi,A,B)&= \frac{1}{2f^2} \int \d^4x (\partial_i \phi+A_i)(\partial^i\phi+A^i) +\frac{\i }{3!(2\pi)}\int \d^4x\epsilon^{ijkl} A_i H_{jkl}\cr
&=\frac{1}{2f^2} \int \d^4x (\partial_i \phi+A_i)(\partial^i\phi+A^i) +\frac{\i }{2\pi}\int A\wedge H \end{align}
is equivalent to the original theory of $\phi$ only.  This follows by performing first the path integral over $B$, after which we can set $A=0$ by a gauge transformation
and we get back to the original starting point in terms of $\phi$.   But we can analyze the path integral in a different way.   First we fix $\phi=0$ by a gauge transformation.
The integral over $A$ is then a Gaussian:
\begin{align}\label{gaussa}\int DA &\exp\left(-\frac{1}{2 f^2}\int \d^4x \sqrt g A_i A^i -\frac{\i }{3!(2\pi)}\int \d^4x\epsilon^{ijkl} A_i H_{jkl}\right)\cr &=\exp\left(-\frac{f^2}{2\cdot 3!(2\pi)^2 }\int \d^4x \sqrt g
H_{ijk}H^{ijk}\right). \end{align}
What remains is then a theory of the two-form field $B$ with an action $\frac{f^2}{2\cdot 3!(2\pi)^2} \int \d^4x \sqrt g
H_{ijk}H^{ijk}$.     This action has the standard form claimed
 in eqn. (\ref{dualityz}), with the expected relation $h=2\pi/f$ between the couplings.   Thus we have deduced the duality betwen $\phi$ and $B$.

\subsection{Transformation of Local Operators}\label{transform}

Now we would like to determine how local operators transform under this duality.   First of all, $\phi(x)$ itself is not a valid local operator,
as it is not invariant under $\phi\to \phi+2\pi$.   A valid operator is instead $\partial_i\phi(x)$, the derivative of $\phi$.  We will begin by determining how this
operator transforms.

 The first step to transform $\partial_i\phi(x)$ under
the duality is to promote it to a gauge-invariant operator in the extended theory of the fields $\phi,A,B$.   This is easily done; a gauge-invariant extension of
$\partial_i\phi(x)$ is $\O_i(x)=\partial_i\phi(x)+A_i(x)$.   Clearly, in the extended theory, if we perform first the path integral over $B$ to impose that $A=0$ up to a gauge transformation,
and then go to the gauge $A=0$, $\O_i(x)$ will reduce to $\partial_i \phi(x)$.   So an insertion of $\O_i(x)$ in the extended theory is equivalent to an insertion of $\partial_i\phi(x)$ in the
original theory of $\phi$ only.   

On the other hand, we can go to the gauge $\phi=0$, in which case $\O_i(x)$ reduces to $A_i(x)$.   Then we integrate over $A$ to find an equivalent operator in the theory of the $B$-field:
\begin{align}\label{nussa}\int DA\,\,\, &A_i(x)\, \exp\left(-\frac{1}{2 f^2}\int \d^4x \sqrt g A_i A^i -\frac{\i }{3!(2\pi)}\int \d^4x\epsilon^{ijkl} A_i H_{jkl}\right)\cr &=-\frac{\i f^2}{3!(2\pi)}\epsilon_{ijkl}
H^{jkl} \exp\left(-\frac{f^2}{2(2\pi)^2 }\int \d^4x \sqrt g
H_{ijk}H^{ijk}\right). \end{align}
On the right hand side, we find an insertion of the operator $-\frac{\i f^2}{3!(2\pi)}\epsilon_{ijkl}
H^{jkl} $.   This gives us the mapping between operators on the two sides of the duality:
\be\label{goodlook} \partial_i\phi\leftrightarrow-\frac{ \i f^2}{3!(2\pi)}\epsilon_{ijkl}
H^{jkl}.\ee   (The overall sign in this relation depends on an arbitrary sign choice in the coupling of $A$ and $B$ in eqn. (\ref{zingo}).)
In this derivation, we considered a single insertion of  $\partial_i\phi(x)$, but the derivation would proceed in the same way with any number of such insertions.    

 The derivation shows that in Euclidean signature,\footnote{In a similar derivation in Lorentz signature, 
the scalar kinetic energy is $\frac{\i}{2f^2}\int\d^4x \sqrt g\partial_i\phi\partial^i\phi$, with an extra factor of $\i$.  As a result,  the relation analogous to eqn. (\ref{goodlook}) has a real coefficient.}
there is a factor of $-\i$ in this mapping between operators.     Correlation functions will be invariant under the substitution (\ref{goodlook}) with the
factor of $-\i$.   This operator substitution is not a transformation between classical fields, and the factor of $-\i$ does not mean that $B$ or $\phi$ should be considered to be imaginary in Lorentz
signature.   Indeed, the derivation was carried
out in terms of path integrals over real  fields $\phi$, $A$, and $B$. 

We will illustrate in a special case the fact that a factor of $\i$ is really needed to get a correct relationship between correlation functions.   Take $M=\R^4$ with metric $\sum_{i=1}^4\d x_i^2$.
Abbreviate the points $(x_1,x_2,x_3,x_4)=(\pm T,0,0,0)$ as $(\pm T,\vec 0)$.   A special case of the relation among correlation functions that follows from the equivalence (\ref{goodlook}) is
\be\label{zongo}\biggl\langle \partial_1 \phi(T,\vec 0)\partial_1 \phi(-T,\vec 0)\biggr\rangle=-\left(\frac{f^2}{2\pi}\right)^2 \biggl\langle H_{234}(T,\vec 0)H_{234}(-T,\vec 0)\biggr\rangle. \ee
The minus sign arose by squaring the factor of $-\i$ in the relation between $\d\phi$ and $H$.
We want to verify the need for this minus sign.  The operators $H_{234}(\pm T,\vec 0)$ are exchanged
by a reflection $(x_1,x_2,x_3,x_4)\leftrightarrow (-x_1,x_2,x_3,x_4)$, so reflection positivity of quantum field theory tells us that $\bigl\langle H_{234}(T,\vec 0) H_{234}(-T,\vec 0)\bigr\rangle>0$.
But the same reflection exchanges $\partial_1 \phi(T,\vec 0)$ with $-\partial_1 \phi(-T,\vec 0)$, so reflection positivity implies that $\bigl\langle \partial_1 \phi(T,\vec 0)\partial_1 \phi(-T,\vec 0)\bigr\rangle<0$.
This confirms that the factor of $-\i$ in eqn. (\ref{goodlook}) is needed in order to match Euclidean correlation functions of the two theories.  But clearly this has nothing to do with either
$B$ or $\phi$ being imaginary or complex in Euclidean signature.

\begin{figure}
 \begin{center}
   \includegraphics[width=2.5in]{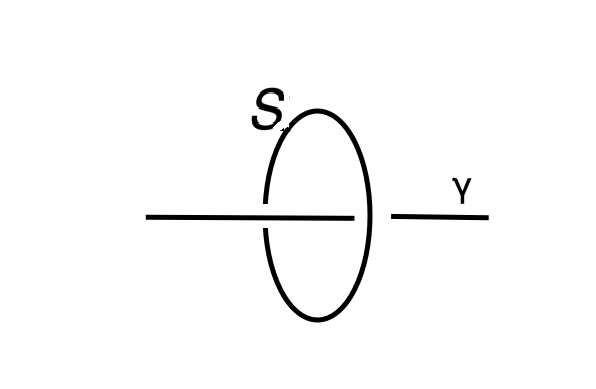}
 \end{center}
\caption{\footnotesize  The two-sphere $S$ (shown here as a circle) links the curve $\gamma$.}\label{Linking}
\end{figure}

It is a little more challenging to understand the transformation of local operators that are not invariant under the symmetry that shifts $\phi$ by a constant.   Such an operator is
$e^{\i \phi(x)}$, or more generally $e^{\i m\phi(x)}$, for an integer $m$.   Before beginning a derivation, let us anticipate what is 
going to happen.   Recall electric-magnetic duality in four dimensions.   This is a relation between a $U(1)$ gauge field $A$ and a dual $U(1)$ 
gauge field $\t A$ that exchanges electricity and magnetism
 and in particular exchanges Wilson operators of $A$ with 't Hooft operators of $\t A$.   For a loop $\gamma$ in spacetime, the Wilson operator is simply a holonomy
$\exp(\i m\oint_\gamma A)$, roughly analogous to $e^{\i m \phi(x)}$ for the scalar field.  By contrast, the 't Hooft operator cannot be described by a function of the dual gauge field $\t A$;
rather, it creates a singularity in $\t A$ along $\gamma$. The singularity is as follows.   Let $S$ be a two-sphere that 
``links'' around $\gamma$ (fig. \ref{Linking}).   Then the recipe  to incorporate a charge $m$
't Hooft operator in the $\t A$ gauge theory is as follows   \cite{Kapustin}:   we use the standard action for $\t A$ as in the absence of the 't Hooft operator, but we evaluate this action and the associated
path integral in a space of fields such that $\t A$ has  a singularity along $\gamma$, characterized by
\be\label{tondo}\int_S \t F =2\pi m,~~~~\t F=\d\t A. \ee

Duality for the operator $e^{\i m \phi(x)}$ in the scalar field theory is similar.   The dual is a ``magnetic'' operator that we will call $K_m(x)$. It can be  described by saying that it
 creates the following type of singularity in the $B$-field.    If $S$ is a three-sphere that wraps around the point $x$ (for example, $S$ may be the boundary of
a ball centered at $x$) then
\be\label{indicot}\int_S H=2\pi m \ee
Note that the relation (\ref{indicot}) is not consistent with the usual Bianchi identity $\d H=0$, but it is consistent with a modified Bianchi identity
with a ``magnetic source'' at $x$:
\be\label{nindicot}\d H= 2\pi m \delta_x.\ee
Here $\delta_x$ is a four-form delta function\footnote{Thus, if the point $x$ is defined by $x_1=x_2=x_3=x_4=0$, then 
$\delta_x= \delta(x_1)\delta(x_2)\delta(x_3)\delta(x_4)\d x_1\d x_2\d x_3 \d x_4$.} supported at the point $x$.   
In the presence of the operator $K_m(x)$, the $B$-field is only defined away from the point $x$, and has the ``magnetic''  singularity just described at the point $x$.
We then study the action and the path integral of the $B$-field in the usual way, except that the $B$-field is constrained to have this magnetic
singularity.  

For the derivation, naively the first step is to promote the operator $W_m=e^{\i m \phi(x)}$ to a gauge-invariant operator of the extended system with $\phi,A,B$.   However,
it is impossible to do this (at least on a compact manifold $M$), since $W_m$ is not invariant under the symmetry that is being gauged.    What we can do instead is to consider
a product of operators $e^{\i m \phi(x)} e^{-\i m\phi(y)}$, with disjoint points $x,y$.   We will show that the dual of this product is $K_m(x) K_{-m}(y)$.   Essentially by cluster decomposition,
the duality relation for an individual operator follows:  $K_m(x)$ is dual to $e^{\i m \phi(x)}$.

If $\gamma$ is a path from $y$ to $x$, then  a gauge-invariant extension of $e^{\i m \phi(x)} e^{-\i m \phi(y)}$ can be made by including a Wilson operator supported on $\gamma$:
\be\label{pindo}\h W_m(x,y)=e^{\i m \phi(x)}\exp\left(\i m\int_\gamma A\right) e^{-\i m \phi(y)}.\ee
The choice of $\gamma$ will not matter, because as usual we can first integrate over $B$ to learn that $A=0$ up to a gauge transformation, and then go to the gauge $A=0$.  
It is useful to describe $\h W_m(x,y)$  in a slightly different way.   Let $\Theta_\gamma$ be a three-form delta function supported on $\gamma$, normalized so that its integral over
a slice transverse to $\gamma $ is 1.   If $\gamma$ is parametrized locally by $x_1$ and defined locally by equations $x_2=x_3=x_4=0$ (where $\d x_1\d x_2\d x_3\d x_4$ defines
the orientation of $M$), then in those coordinates $\Theta_\gamma=\delta(x_2)\delta(x_3)\delta(x_4)
\d x^2\d x^3\d x^4$.    We can then alternatively write
\be\label{lindo} \h W_m(x,y)= e^{\i m \phi(x)} \exp\biggl(\i m\int_M A\wedge \Theta_\gamma\biggr)e^{-\i m \phi(y)}.\ee
We may observe that 
\be\label{showme}\d\Theta_\gamma =-\delta_x+\delta_y,\ee
where $\delta_x $ and $\delta_y$ are four-form delta functions supported at $x$ and $y$, respectively.   This is the only property of $\Theta_\gamma$ that is needed in order to
verify gauge-invariance of $\h W_m(x,y)$ as defined in eqn. (\ref{showme}).   So in fact, we can replace $m\Theta_\gamma$ with $\Theta$, where $\Theta$ is any three-form 
that satisfies
\be\label{knowme}\d\Theta=m(-\delta_x+\delta_y). \ee
So a more general choice of  $\h W_m(x,y) $ is
\be\label{lowme}\h W_m(x,y)= e^{\i m \phi(x)} \exp\left(\i \int_M A\wedge \Theta\right)e^{-\i m \phi(y)}.\ee

As usual, the choice of $\Theta$ does not matter, since we can integrate first over $B$ and reduce to $A=0$.
But instead, also as usual, we can go to the gauge $\phi=0$ and integrate over $A$.   Compared to the integral that was involved in deriving the duality, the integrand now has
an extra factor  $\exp(\i \int_M A\wedge \Theta)$.   The effect of this is just to replace $H$ by $H-2\pi \Theta$.   Indeed,  eqn. (\ref{gaussa}) is replaced with
\begin{align}\label{naussa}\int DA &\exp\left(-\frac{1}{2 f^2}\int \d^4x \sqrt g A_i A^i -
\frac{\i }{3!(2\pi)}\int \d^4x\epsilon^{ijkl} A_i (H_{jkl}-2\pi\Theta_{jkl})\right)\cr &=\exp\left(-\frac{f^2}{2(2\pi)^2 3! }\int \d^4x \sqrt g
(H_{ijk}-2\pi\Theta_{ijk})(H^{ijk}-2\pi\Theta^{ijk})\right). \end{align}  In the exponent, we see the original action but with $H$ replaced by $H-2\pi\Theta$.
At this point, we may as well redefine $H-2\pi\Theta$ as $H$.   The action for the shifted $H$-field then takes the standard form $\frac{f^2}{2(2\pi^2)3!}\int \d^4x \sqrt g H_{ijk} H^{ijk}$, but the newly
defined $H$ obeys a modified Bianchi identity:
\be\label{modbia}\d H=2\pi m\left(\delta_x-\delta_y\right). \ee
This modification of the Bianchi identity corresponds by definition to insertion of the operator $K_m$ at $x$ and $K_{-m}$ at $y$.  
Thus, we have justified the claim that the dual of $e^{\i m \phi(x)}$ is the magnetic or 't Hooft-like operator $K_m(x)$.

\subsection{Dual Description of Vacua}\label{dualvac}

So far, we have considered a compact spacetime manifold $M$.   Now we will consider instead the case that $M$ is, for example, asymptotically flat.  In that case, in the theory of the massless
scalar field $\phi$, 
we have the option to choose an arbitrary angle $\alpha$ and to perform a path integral with $\phi\to\alpha$ at infinity.   This amounts to studying the theory in a family of quantum vacuum
states parametrized by $\alpha$.   How can one describe the parameter $\alpha$ in the two-form language?  First we will explain in the two-form language
how to incorporate the parameter $\alpha$ in Euclidean correlation functions.    Then we will go to Lorentz signature and explain how to construct an $\alpha$-dependent family
of quantum vacuum states in the two-form language.   

We will denote an expectation value computed with the condition that $\phi\to\alpha$ at infinity  as $\left\langle \cdot \right\rangle_\alpha$.
For any integers $m_i$ and points $x_i$,
\be\label{tonogo} \left\langle \prod_{i=1}^s e^{\i m_i \phi(x_i)} \right\rangle_{\negthinspace\negthinspace\alpha} = C e^{\i \alpha\sum_i m_i}, \ee
where $C$ depends on the points $x_i$ and the integers $m_i$, but not on $\alpha$.   Indeed, this follows from the underlying spontaneously broken symmetry: adding to $\phi$
the constant $\alpha$  multiplies the product of operators $\prod_{i=1}^s e^{\i m_i \phi(x_i)}$  by $e^{\i\alpha\sum_i m_i}$.
To reproduce this result in terms of $B$, we need to find a way to give a phase $e^{\i\alpha\sum_i m_i}$ to any expectation value $\left\langle \prod_{i=1}^s K_{m_i}(x_i)\right\rangle$.
There is a straightforward recipe.   In the presence of the operator product $\prod_{i=1}^s K_{m_i}(x_i)$, the total flux at infinity of the three-form $H$ is
\be\label{nillow}\int_{\partial M} H = 2\pi\sum_i m_i. \ee   
So to get a phase $\exp(\i \alpha \sum_i m_i)$, we just need to include in the path integral an extra factor 
\be\label{ilow}\exp\left(\frac{\i \alpha}{2\pi}\int_{\partial M} H \right). \ee

This gives a recipe to reproduce $\alpha$-dependent Euclidean correlation functions in the two-form language.  But in Minkowski space, does the $B$-field theory have an $\alpha$-dependent
family of vacuum states, as the scalar field theory does?   Indeed, it does.
Pick a time-zero slice $W\cong \R^3$ in Minkowski space.   A quantum state of the $B$-field is a functional $\Psi(B)$ of the $B$-field restricted to $W$.
In particular, the vacuum at $\alpha=0$ is described by such a wavefunctional $\Psi_0(B)$.   
  Then a vacuum with a general value of $\alpha$ can be described by the wavefunctional 
\be\label{ohnot} \Psi_\alpha(B)=\exp\left(-\i \frac{\alpha}{2\pi}\int_W H\right)\Psi_0(B). \ee
One way to justify this statement is simply to note that $\frac{1}{2\pi}\int_W H$ is dual, according to eqn. (\ref{goodlook}), to the canonical momentum $\frac{1}{f^2}\int \d^3x \dot\phi$
that generates a constant shift of $\phi$.   
More explicitly, we can show that $\la\Psi_\alpha|K_m(x)|\Psi_\alpha\ra$ is proportional to $e^{im\alpha}$ (times a constant independent of $\alpha$).
Indeed from the definition of $\Psi_\alpha$ it follows that
\be\label{hnot}\la\Psi_\alpha|K_m(x)|\Psi_\alpha\ra =\la \Psi_0| U_\alpha^\dagger K_m(x) U_\alpha| \Psi_0\ra, \ee
where $U_\alpha$ is the unitary operator $U_\alpha=\exp\left(-\i\frac{\alpha}{2\pi}\int_W H\right)$.  Since $K_m(x)$ increases the flux of $H$ by $m$ units, we have
$U_\alpha^\dagger K_m(x) U_\alpha=e^{\i m\alpha} K_m(x)$, implying that $\la\Psi_\alpha|K_m(x)|\Psi_\alpha\ra $ is proportional to $e^{i  m\alpha}$, as desired.

Going back to Euclidean signature, in  studying the axion wormhole, we will be interested in a case in which $M$ has two asymptotically flat ends.   In that case,
separate $\alpha$ parameters can be chosen at the two ends.  In  the two-form language, these parameters can be incorporated by including a separate factor of the form of
eqn. (\ref{ilow}) at each end.

\section{The Wormhole Solution}{\label{wormhole}

\subsection{The Solution}

In this section, we will review the axion wormhole solution of \cite{GS}, and then describe in this geometry the path integral over $\phi$ or $B$ and the duality between them.
The main point is to show that this duality is not a simple map between classical fields, but involves a Poisson resummation.   In the process it will become clear that, in contrast to $B$,
$\phi$ cannot be treated semiclassically in the wormhole throat.   

First of all, the wormhole spacetime, which we will call\footnote{A generic four-manifold will be called $M$, while $X$ is specifically the wormhole spacetime.}  $X$,  is topologically $\R\times S^3$, where $\R$ is parametrized by a variable $r$ that has the full range $-\infty<r<\infty$.     The metric is
\be\label{yogo} \d s^2= \d r^2+a^2(r)\d\Omega^2,\ee
where $\d\Omega^2$ is the metric of a round three-sphere of unit radius.   In particular, this geometry has $SO(4)$ rotational symmetry.   The function $a(r)$ will behave
as $a(r)\sim |r|$ for $r\to \pm\infty$, and therefore the wormhole has two asymptotically flat ends, for $r\to +\infty$ and $r\to -\infty$, as sketched in fig. \ref{Tube}(a).    We orient
$X$ by the four-form $\d r\wedge \Omega$, where $\Omega$ defines the orientation of $S^3$.

The metric ansatz (\ref{yogo}) is a Euclidean version of a standard FLRW ansatz in cosmology, and accordingly the Einstein equations imply a Euclidean version of
the FLRW equation:
\be\label{insto} -\left(\frac{a'}{a}\right)^2=\frac{8\pi G}{3}\rho-\frac{1}{a^2}, \ee
where $a'=\frac{\d a}{\d r}$, and $\rho$ is the energy density.

We choose $H$ to be a multiple of the Levi-Civita form of $S^3$, normalized so that the total flux is $m$:
\be\label{winsto}\frac{1}{2\pi}\int_{S^3} H = m . \ee
Since the volume of $S^3$ is $V(r)=2\pi^2 a(r)^3$, the energy density computed from the action (\ref{action}) is $\frac{1}{2h^2} \frac{(2\pi m)^2}{V(r)^2}=\frac{f^2m^2}{2(2\pi^2)^2 a(r)^6}$.
The FLRW equation therefore becomes
\be\label{ninto} -\left(\frac{a'}{a}\right)^2=\frac{Gf^2m^2}{3\pi^3 a(r)^6}-\frac{1}{a^2},\ee
implying that
\be\label{winto}\d r =\pm \frac{\d a}{\sqrt{1- \frac{a_0^4}{a^4}}},\ee
with
\be\label{pinto}a_0=\left(\frac{Gf^2 m^2}{3\pi^3}\right)^{1/4}. \ee
So
\be\label{linto}r=\pm \int_{a_0}^a\frac{\d\t a}{\sqrt{1-\frac{a_0^4}{\t a^4}}}. \ee
The two branches with $r\geq  0$ and $r\leq 0$ meet smoothly at $r=0$, $a=a_0$.  
For $r\to \pm\infty$, we have 
\be\label{pintok} a\sim|r|+{\mathrm{constant}},\ee
as asserted earlier.  

At any given $r$, this geometry consists of a three-sphere of radius $a$.   The minimum value $a=a_0$ is reached at $r=0$, deep inside the wormhole ``throat.''
One can think of $a_0$ as the radius of the throat.     For
a reliable semiclassical solution, this should be larger than the Planck length $G^{1/2}$.   The condition $a_0\gg G^{1/2}$ is equivalent to
\be\label{intelz} f^2m^2 \gg G, \ee meaning that $f$ and/or $m$ should be sufficiently large.
For a reliable semiclassical solution, one would also like the field strength $H$ to be small in magnitude compared to $G^{-3/2}$.   As the maximum value of $H$ is
$H\sim 1/a_0^3$, achieved in the wormhole throat, the condition $H\ll G^{-3/2}$ gives again $a_0\gg G^{1/2}$.   

\subsection{A Practice Exercise}\label{practice}

Provided the condition $a_0\gg G^{1/2}$ is satisfied, the classical wormhole solution of the two-form theory is reliable semiclassically, and in principle quantum corrections can
be computed by a weakly coupled perturbation theory.  (Renormalization is needed, of course, and in practice, some thorny issues come up; a recent reference is \cite{MarolfM}.)   We would like to understand
a dual description of the wormhole spacetime in terms of the scalar.   This is less straightforward, since the wormhole does not have a semiclassical description in terms of the
scalar, as observed in  \cite{GS} and as explained in the introduction to this article.   In section \ref{effac}, we will explain what sort of path integral describes
the wormhole in the scalar language, and how it can be used to compute the effective action associated to the wormhole.

Here we will consider a simpler practice exercise.   
We will consider
the metric of $X$ to be fixed, with some specific function $a(r)$ in eqn. (\ref{yogo}).   In computing the partition function in the two-form language, we will have to sum
over the flux $m$ defined in eqn. (\ref{winsto}).    We will do that sum keeping the metric of $X$ fixed, ignoring the fact that
 imposing the Einstein equations would determine the metric in terms of $m$,
via the formulas just presented.    The goal of the exercise is to illustrate the fact that the duality between a two-form and a scalar is not a simple map of classical fields to classical fields,
but involves a Poisson resummation in the sum over fluxes.   

The path integral over either the two-form or the scalar is the product of a sum over fluxes times a one-loop determinant of small fluctuations.  The
full path integral including the determinants has been analyzed in   \cite{DMW}, and will be discussed in Appendix \ref{zerograv}.  Here we simply compare the sums over fluxes.

In the two-form language, we have to sum over the integer $m=\frac{1}{2\pi}\int_{S^3} H. $   For given $m$, the value of the action is
\be\label{zac}I_m=\frac{1}{2h^2}\int_{-\infty}^\infty\frac{\d r}{a^3}\frac{(2\pi m)^2}{2\pi^2} =\frac{f^2}{4\pi^2}\int_{-\infty}^\infty \frac{\d r }{a^3} .\ee
The integral is
\be\label{wac}\int_{-\infty}^\infty\frac{\d r}{a^3} =2\int_{a_0}^\infty \frac{\d a}{a^3}\frac{1}{\sqrt{1-\frac{a_0^4}{a^4}}}=\frac{\pi}{2a_0^2},\ee
so the action is
\be\label{whod} I_m=\frac{f^2m^2}{8\pi a_0^2}.\ee
The sum over fluxes is thus a theta function:
\be\label{thetaB} \Theta_B=\sum_{m=-\infty}^\infty \exp\left(-\frac{f^2m^2}{8\pi a_0^2}\right).\ee

In analyzing the scalar partition function, to begin with, we assume a boundary condition such that $\phi\to 0$ at infinity (mod $2\pi$) on either side of the wormhole.     There is then an integer-invariant: the change in $\phi$ in going from $r=-\infty$ to $r=+\infty$ 
can be an arbitrary integer multiple of $2\pi$.   It is convenient to find a classical solution $\phi_0$ that  vanishes at infinity (mod $2\pi$) at both ends
and jumps by precisely $2\pi$ in passing through the wormhole.  
Then we can write
\be\label{hopwrite}\phi=\varphi+n\phi_0,\ee
where $\varphi$ is a real-valued field and $n$ is an integer.   The path integral over $\phi$ is then the product of a sum over $n$ and a Gaussian integral over $\varphi$.  
The sum over $n$ gives a theta function that we will compare to $\Theta_B$.   The Gaussian integral over $\varphi$ gives a determinant that is discussed in Appendix \ref{zerograv}.   

The relevant classical solution $\phi_0$ is invariant under rotations of the wormhole spacetime and is a function only of $r$.
The equation that it has to obey is then $\frac{\d}{\d r}a^3(r)\frac{\d}{\d r}\phi_0=0$.   The general solution vanishing for $r\to -\infty$ is
$\phi_0(r) = C\int_{-\infty}^r \d \t r/a^3(\t r)$, with a constant $C$.   The integral is the same one that we encountered in eqn. (\ref{wac}).    We set $C=4 a_0^2$ so that the
$\phi_0(\infty)-\phi_0(-\infty)=2\pi$, and find the asymptotic behavior of $\phi_0$:
\be\label{asbe}\phi_0(r)\approx \begin{cases} 2\pi -\frac{2 a_0^2}{r^2} & r\to+\infty\cr
                                                                                \frac{2 a_0^2}{r^2} & r\to -\infty.\end{cases}\ee

Now we want to evaluate the action $I(\phi)=\frac{1}{2f^2}\int \d^4 x \sqrt g \partial_i\phi \partial^i\phi$, with $\phi=\varphi+n\phi_0$.   Because $\phi_0$ satisfies the classical equation of motion
and $\varphi$ vanishes at infinity, a simple integration by parts shows that in the evaluation $I(\phi)$, there is no cross term between $\varphi$ and $n\phi_0$; the action is
the sum $I(\phi)=I(\varphi)+I(n\phi_0)$.  Using rotation symmetry and integrating by parts, we have
\be\label{nasbe}I(n\phi_0)=\frac{n^2}{2f^2} 2\pi^2\int_{-\infty}^\infty \d r \, a^3(r) (\partial_r\phi_0)^2=\frac{\pi^2n^2}{f^2}\left[ a^3(r) \phi_0\partial_r\phi_0\right]_{-\infty}^{\infty}=\frac{8\pi^3 n^2a_0^2}{f^2}.
\ee
The sum over $n$ is thus another theta function:
\be\label{welpn}\Theta_\phi=\sum_{n=-\infty}^\infty \exp\left(-\frac{8\pi^3 n^2a_0^2}{f^2}\right).\ee

The two theta functions $\Theta_\phi$ and $\Theta_B$ can be related to each other by Poisson resummation.  We have
\begin{align} \label{Poisson}\Theta_\phi=&\sum_{m=-\infty}^\infty \int_{-\infty}^\infty\d n\, e^{-2\pi \i m n}\exp\left(-\frac{8\pi^3 n^2a_0^2}{f^2}\right)\cr =&\frac{f}{2\pi\sqrt 2 a_0} \sum_{m=-\infty}^\infty
 \exp\left(-\frac{f^2m^2}{8\pi a_0^2}\right)=\frac{f}{2\pi\sqrt 2 a_0} \Theta_B. 
\end{align}
Thus, the two theta functions actually agree, up to the elementary factor $f/2\pi\sqrt 2 a_0$.   To understand this factor involves a counting of zero-modes (including zero-modes of
ghosts and ghosts for ghosts) as well as a gravitational counterterm first identified in \cite{DMW}.   See Appendix \ref{zerograv}.

It is straightforward to generalize this analysis to the case that $\phi$ approaches specified angles $\alpha_+$ and $\alpha_-$ for $r\to \pm\infty$.    
Setting $\alpha=\alpha_+-\alpha_-$, now we expand $\phi=\varphi+(n+\alpha/2\pi)\phi_0+\alpha_-$.   The constant $\alpha_-$ does not affect the path integral; it can be removed
using the shift symmetry of $\phi$.
The path integral for $\phi$ is modified only by shifting the exponents in the theta function:
\be\label{zelpn}\Theta^{(\alpha)}_{\phi}=\sum_{n=-\infty}^\infty \exp\left(-\frac{8\pi^3 (n+\alpha/2\pi)^2a_0^2}{f^2}\right).\ee
Poisson resummation now gives a slightly different answer: 
\begin{align} \Theta_\phi^{(\alpha)}=&\sum_{m=-\infty}^\infty \int_{-\infty}^\infty\d n\, e^{-2\pi \i m n}\exp\left(-\frac{8\pi^3 (n+\alpha/2\pi)^2a_0^2}{f^2}\right)\cr =&\frac{f}{2\pi\sqrt 2 a_0} \sum_{m=-\infty}^\infty
 \exp\left(-\frac{f^2m^2}{8\pi a_0^2}\right)e^{\i m\alpha}. 
\end{align}
Here we see the generalization of $\Theta_B$ to include $\alpha$:
\be\label{thetba}\Theta_B^{(\alpha)}= \sum_{m=-\infty}^\infty
 \exp\left(-\frac{f^2m^2}{8\pi a_0^2}\right)e^{\i m\alpha}. 
 \ee
 The flux parameter $m$ is still an integer, as it must be according to Dirac quantization.   But a contribution with a given flux $2\pi m$ at $r=\infty$ is weighted by
 an extra factor $e^{\i m\alpha}$.   The origin of this factor was explained in section \ref{dualvac}.   To describe in terms of the $B$-field a vacuum in which, in the description by
 a scalar field, 
 $\phi\to\alpha$ for $r\to\infty$,   one has to include in the path integral a factor $\exp\left(\i \frac{\alpha}{2\pi}\int_{S_\infty} H\right), $
 where $S_\infty$ is a three-sphere at $r\to\infty$.    In the present context, this factor becomes $e^{\i m\alpha}$.

\section{The Effective Action}\label{effac}

Consider an axion wormhole connecting two asymptotically flat worlds.      From a macroscopic point of view,  the wormhole spacetime consists of
 two Euclidean spaces $\R^4$ and $\t \R^4$ that are glued together by the wormhole.   The gluing is of course not really a local operation, but 
at length scales much greater than the wormhole size $a_0$, it looks effectively local, and
  can be viewed simply as an identification of a point $x\in \R^4$ with a point $y\in \t\R^4$.
  The points $x,y$ are arbitrary and are moduli of the wormhole.\footnote{The axion wormhole also has rotational moduli that express the fact that it is possible
  to rotate one end of the wormhole spacetime relative to the other.   Integrating over these moduli has the effect of restricting the sum in eqn. (\ref{ibox}) to a sum over rotationally
  invariant operators only.    This will not play an important role in the present article, because the operator of lowest dimension that could appear in eqn. (\ref{ibox}) is anyway
  rotationally invariant.}   To construct the effective action, we have to integrate
over these moduli.   This  naturally leads to a bilocal effective action:
\be\label{ibox} I_\eff=\sum_i c_i \int_{\R^4\times\t\R^4}\d^4x \,\d^4 y \,\,\O_i(x) \cdot \t \O_i(y). \ee
The idea behind this formula is that, by a sort of operator-state correspondence,  a quantum state $\Psi_i$ propagating ``through the wormhole'' from $\t \R^4$ to $\R^4$ will  produce an effect
that on $\R^4$ can be simulated by an insertion of an operator $\O_i$, while an observer on $\t \R^4$ would say that a conjugate state $\t\Psi_i$ propagating through the wormhole from $\R^4$ to
$\t \R^4$  can be simulated by an insertion of  a conjugate operator $\t\O_i$,   The constants $c_i$
in the effective action are $c$-numbers that measure the amplitude for the quantum state to propagate through the wormhole.

The sum in eqn. (\ref{ibox}) is an infinite sum, but from a long distance point of view it will be rapidly convergent, dominated by the operator or operators of lowest dimension.

In the case of the axion wormhole, it is straightforward to implement this program in the two-form language.   By definition, in the wormhole spacetime, there are $m$ units of $H$-flux
measured at infinity on $\R^4$, and $-m$ units measured at infinity on $\t\R^4$.   So the relevant operators $\O_i$ create $m$ units of $H$-flux, while the $\t \O_i$ create
$-m$ units.   The lowest dimension operators with this property are the 't Hooft-like operators $K_m(x)$ and $K_{-m}(y)$ described in section \ref{transform}.   Other possible
operators are simply the product of $K_m(x)$ or $K_{-m}(y)$ with a polynomial in $H$ and its derivatives.   So at long distances, the dominant term in the effective action  is
\be\label{zimbox}I_\eff\approx c \int_{\R^4\times \t\R^4}\d^4x \d^4y \,K_m(x) K_{-m}(y),\ee   with some constant $c$,  where the symbol $\approx$ means that the relation holds asymptotically at large distances.

Duality between the two-form and the scalar tells us then what the effective action must be in the description by a scalar field.  Since operators $K_{\pm m}$ are dual  to
$e^{\pm \i m \phi}$, the effective action in the scalar language must be dominated at long distances by
\be\label{limbo} I_\eff\approx c \int_{\R^4\times \t\R^4} \d^4 x \d^4y \, e^{\i m \phi(x)} e^{-\i m \phi(y)}. \ee
We would like to do a direct calculation in the scalar language to exhibit this result.    This is subtle because, as is hopefully clear from the practice exercise in
section \ref{practice}, the wormhole cannot be described in the scalar language by a single classical field, only by a sum over classical fields.

Let us first explain what the strategy will be to identify the effective action.   Let $\la ~\ra_X$ denote a path integral computed in the wormhole spacetime, and let 
$\la~\ra_{\R^4}$ or $\la ~\ra_{\t\R^4}$ denote a path integral computed on $\R^4$ or $\t\R^4$.   Let $\O(z)$ be a test operator supported at a point $z\in \R^4$
that is far from the wormhole mouth.   Then eqn. (\ref{limbo}) predicts that
\be\label{himbo} \la \O(z)\ra_X\approx c \la \O(z) e^{\i m \phi(x)}\ra_{\R^4} \la e^{-\i m \phi(y)}\ra_{\t \R^4}.\ee
We expect similar formulas for a case with several operators $\O_i(z_i)$ inserted on $\R^4$ at points  distant from $x$, and several additional operators
$\t\O_j(z_j)  $ inserted on $\t\R^4$ at points distant from $y$:
\be\label{limbot}\la \prod_i \O_i(z_i)\prod_j\t\O_j(z_j)\ra_X \approx  c \la \prod_i \O_i(z_i)e^{\i m \phi(x)}\ra_{\R^4}\la \prod_j\t\O_j(z_j) e^{-\i m \phi(y)}\ra_{\t \R^4}. \ee
These relations are expected to hold regardless of the boundary conditions assumed for $r\to\pm\infty$, but in testing them, we will assume to begin with that $\phi\to 0$ at $r\to \pm\infty$.   
This means, for example, that eqn. (\ref{himbo}) simplifies to 
\be\label{himbox} \la \O(z)\ra_X\approx c \la \O(z) e^{\i m \phi(x)}\ra_{\R^4} ,\ee
since $ \la e^{-\i m \phi(y)}\ra_{\t \R^4}=1$ if $\phi\to 0$ at infinity.

For a choice of test operator, we cannot take $\O(z)=\phi(z)$, which is not invariant under $\phi\to \phi+2\pi$.   Instead we will use test operators  $\O_i(z)=\partial_i\phi(z)$.

The basic idea of the derivation is to use the extended theory of fields $\phi,A,B$, but with a constraint on the value of the $H$-flux.   This constraint could be incorporated with a simple
factor
\be\label{toddo} \int_0^1\d x e^{\i x(\int_S H -2\pi m)}.\ee
Here $S$ is a three-sphere at an arbitrary value of $r$; the integral equals 1 if the $H$-flux is $2\pi m$ and 0 otherwise.
The value of $r$ is arbitrary here, since the flux does not depend on $r$.   Therefore, we could equivalently average over $r$ with an arbitrary weight.   A convenient choice is to use
the function $\phi_0$ that satisfies $\int_{-\infty}^\infty \d \phi_0=2\pi$ and  impose the constraint via
\be\label{oddo} \int_0^1\d x \,\exp\left(\i x\left(\int_X \frac{\d \phi_0}{2\pi }\wedge H - 2\pi m\right)\right). \ee
Now we consider the extended theory of fields $\phi,A,B$ with the action $I(\phi,A,B)$ (eqn. (\ref{helpme})) and with this additional integral over $x$.   The combined integral, also
with insertion of a test operator, becomes
\be\label{loddo}\int_0^1\d x \int D\phi\,DA\,DB\,e^{-I(\phi,A,B)}  \,\exp\left(\i x\left(\int_X \frac{\d \phi_0}{2\pi }\wedge H - 2\pi m\right)\right)\,\O_i(\phi,A). \ee
If we go to the gauge $\phi=0$ and integrate out $A$, we get the theory of $B$ only with its standard action and a constraint restricting to $\int_S H=2\pi m$ (and a dual version of the
test operator).
Since $m$ is fixed, after coupling to gravity, a definite semi-classical spacetime emerges.
(This can be compared to the practice problem studied in section \ref{practice}, where all values of $m$ were allowed and the gravitational back-reaction, which depends strongly
on $m$, was ignored.)   We will choose the test operator to be  $\O_i(\phi,A)=\partial_i\phi(z)+A_i(z)$, which, as in section \ref{transform}, is the gauge-invariant extension of $\partial_i\phi$.
So this integral will compute $\la\partial_i\phi(z)\ra_X$.

Instead of integrating first over $\phi$ and $A$, we can integrate over $B$ and $A$ to find what is the path integral over $\phi$ that is dual to a wormhole of definite flux.
This is easily done. The integral over $B$ is the same as before, but with $A$ replaced by $A-x\d\phi_0$.   So  rather than the integral over $B$ giving a delta function setting $A=0$ up to a gauge transformation, as in section \ref{derivation}, now we get a delta
function setting $A=x\d\phi_0$ up to a gauge transformation.  After going to the gauge $A=x\d\phi_0$ and integrating over $A$, we arrive at a path integral for $\phi$ only
that is dual to a wormhole of definite $m$, and with a test operator included:
\be\label{ploddo}\int_0^1\d x \,e^{-2\pi \i x m} \int\D \phi \exp\left(-\frac{1}{2f^2}\int \d^4x \left(\partial_i \phi+ x \partial_i\phi_0)(\partial^i\phi+x\partial^i \phi^0)\right)\right) \left(\partial_i\phi(z)+x\partial_i\phi_0(z)\right). \ee

What we {\it cannot} now usefully do is to redefine $\phi+x\phi_0\to \phi$  with the goal of decoupling the integrals over $x$ and $\phi$.   Since $x\phi_0$ does not
vanish at infinity, this would not really decouple $x$ and $\phi$; it would hide the coupling between them in the required asymptotic behavior of $\phi$.   What
we {\it can} usefully do is to make the same expansion $\phi=\varphi+n\phi_0$ as in eqn. (\ref{hopwrite}), where $\varphi$ is a real-valued field that vanishes at infinity and $n$ is an integer.
Upon doing this, we find that the integer $n$ and the  variable $x$
appear only in the combination $n+x$, so we can combine $n$ and $x$ to a real variable $n$ that is integrated over the whole real line.   We thus arrive at
\be\label{loddox}\int_{-\infty}^\infty \d n \int D\varphi\, e^{-2\pi \i n m} \exp\left(-\frac{1}{2f^2}\int\d^4x\sqrt g\left(n^2 \partial_i\phi_0\partial^i\phi^0+\partial_i\varphi\partial^i\varphi\right)\right)
\cdot \left(\partial_i\varphi (z)+ n\partial_i \phi_0(z)\right).\ee
As in section \ref{practice}, we have used the fact that in the action, there is no cross term between $\varphi$ and $\phi_0$.     In the test operator $\partial_i\varphi (z)+ n\partial_i \phi_0(z),$
we can drop the term proportional to $\varphi$, as it is an odd function of $\varphi$, while the action is even in $\varphi$.   The integral over $\varphi$ is then a decoupled Gaussian
integral, equal to a constant $c_0$, which will be one factor in the constant $c$  that appears in eqn. (\ref{limbo}).   At this stage, the integral (\ref{loddo})  reduces
to an integral just over $n$, giving a formula for $\la\partial_i\phi(z)\ra_X$ which we make more explicit using eqn. (\ref{nasbe}):
\begin{align}\label{zoddo}\la\partial_i\phi(z)\ra_X=&c_0\int_{-\infty}^\infty \d n e^{-2\pi \i n m}\left( n\partial_i\phi_0(z)\right)\exp\left(-\frac{8\pi^3 n^2 a_0^2}{f^2}\right)\cr
= &c \frac{-\i f^2 m\partial_i\phi_0(z)}{8\pi^2 a_0^2},\end{align}
with
\be\label{tellnot} c=c_0 \frac{f}{2\pi\sqrt 2 a_0} \exp\left(-\frac{m^2 f^2}{8\pi a_0^2}\right). \ee
Here $c$ is the value of the path integral without the insertion of the test operator $\partial_i\phi(z)$.   (The exponential factor has a simple interpretation:
 $\frac{m^2 f^2}{8\pi a_0^2}$ is just the wormhole action 
 in the original description by the $B$-field.)     So the claim (\ref{himbox}), which we are aiming to test, becomes the assertion that if $z$ is far from the wormhole, then
\be\label{zelcot} \frac{-\i m f^2 \partial_i\phi_0(z)}{8\pi^2 a_0^2}\overset{?}{=}\la \partial_i\phi(z) e^{\i m\phi(x)}\ra_{\R^4}= \i m\la\partial_i\phi(z)\phi(x)\ra_{\R^4}.\ee
We have, from eqn. (\ref{asbe}), $\phi_0(z)\sim -2a_0^2/r^2$, so the left hand side of eqn. (\ref{zelcot}) reduces to $\frac{\i f^2 m}{4\pi^2}\partial_i \frac{1}{r^2}$, with a nice
cancellation of factors of $a_0$.   This agrees with the right hand side of eqn. (\ref{zelcot}), since 
in massless free field theory in four dimensions with action $\frac{1}{2f^2}\int \d^4x \partial_i\phi \partial^i\phi$, we have
$\la\partial_i\phi(z)\phi(x)\ra=\frac{f^2}{4\pi^2 }\partial_i\frac{1}{r^2}$, where $r=|z-x|$.  

As seen in eqn. (\ref{zoddo}), the expectation value $\la \partial_i\phi(z)\ra_X$  in the wormhole spacetime $X$ is imaginary.   This has nothing to do with $\phi$ being imaginary;
we performed a path integral over real values of $\phi$.  Rather, the imaginary answer for $\la \partial_i\phi(z)\ra_X$ resulted from the fact that to reproduce in terms of
$\phi$ a path integral in which $H$ has a definite, non-zero flux, we have to integrate over $\phi$ with a complex measure.
The real function $\partial_i\phi(z)$ turned out to have an imaginary expectation value with respect to this complex measure.  In general, expectation values with respect to complex
measures are not very intuitive.

In this analysis, we probed the effective action with a single insertion of a test operator $\partial_i \phi(z)$.   One can make a similar analysis with
insertion  of any number of  test operators $\partial_{i_j }\phi(z_j)$.    The main difference is that when one expresses $\phi$
as $\varphi+n\phi_0$, the $\varphi$ terms in the test operators cannot be dropped as they may be Wick contracted with each other.   This is as expected, since
likewise in a correlation function  $\la \prod_{j=1}^s \partial_{i_j}\phi(z_j) \,e^{\i m\phi(x)}\ra_{\R^4}$ on $\R^4$, some of the operators $\partial_{i_j}\phi(z_j)$ can be contracted
with each other, rather than with factors of $\phi$ coming from the expansion of  $e^{\i m\phi(x)}$.

There is no difficulty to generalize to the case that $\phi$ vanishes at infinity on one side of the wormhole and $\phi$ approaches a constant $\alpha$ on the other side.
The dual of this, in the description by a $B$-field, is to include in the path integral a factor $e^{\i m \alpha}$, where $m$ is the flux of $H/2\pi$ at infinity.   In our present analysis,
$m$ is treated as a constant, a $c$-number, so the path integral is just multiplied by a constant factor $e^{\i m\alpha}$, independent of $\phi$.   The unnormalized
expectation value $\la\partial_i\phi(z)\ra_X$ is multiplied by the same factor, and a similar normalized expectation value  is unchanged.

More generally, suppose that $\phi$ approaches a constant $\alpha$ at infinity on one side of the wormhole, where the $H$-flux is $m$, and another constant $\alpha'$ on the 
other side of the wormhole, where the flux is $-m$.   This simply has the effect of multiplying all amplitudes by a factor $e^{\i m(\alpha-\alpha')}$.

To summarize, although the scalar field cannot be treated semiclassically in the wormhole throat, there is no problem in using  the scalar field theory to compute the effective
action generated by the wormhole.

\appendix

\section{The Action Revisited}\label{direct}

In deriving the duality between a massless scalar and a two-form gauge field in section \ref{derivation}, we made use of a coupling of a $U(1)$ gauge field $A$ and a field $B$ that
is the two-form analog of a $U(1)$ gauge field.  If either $A$ or $B$ is topologically trivial, this coupling has an elementary definition
\be\label{piml} J(A,B)=\frac{1}{2\pi}\int_M A\wedge H=\frac{1}{2\pi}\int_M F\wedge B, \ee
where $F=\d A$ and $H=\d B$ are the field strengths.  
A related fact is that there is a simple formula for the change in $J(A,B)$ under an infinitesimal variation of $A$ and $B$:
\be\label{ziml}\delta J(A,B)=\frac{1}{2\pi}\int_M\left(\delta A\wedge H+F\wedge \delta B\right).\ee
 There is one more important situation in which $J(A,B)$ has a simple definition.  If there is a manifold $Z$ with boundary $M$
such that $A$ and $B$ extend over $Z$, then
\be\label{iml}J(A,B)=\frac{1}{2\pi} \int_Z F\wedge H. \ee
This is independent mod $2\pi$ of the choice of $Z$ and the extensions of $A$ and $B$, because if $Z$ is a manifold without boundary, then $\frac{1}{2\pi}\int_Z F\wedge H$
is a multiple of $2\pi$, by Dirac quantization.

In general, $A$ and $B$ are topologically nontrivial, $Z$ does not exist, and also the formula (\ref{ziml}) for $\delta J$ does not lead to any immediate way to compute
$J$.   The theory of Cheeger-Simons differential characters gives a general way to deal with this issue
and to define $J(A,B)$ for the general case that $A$ is a $p$-form gauge field and $B$ is an $n-p-1$ gauge field in $n$ dimensions.  See \cite{MooreSaxena,HS}.
However, in the particular case at hand, we can define  $J(A,B)$ as follows.

First of all, the obstruction to topological triviality of $A$ is measured by a first Chern class $\x\in H^2(M,\Z)$, and similarly the obstruction to topological triviality of $B$ is measured
by a class $\y\in H^3(M,\Z)$.   At the level of differential forms, $\x$ and $\y$ are described respectively by $\frac{F}{2\pi}$ and $\frac{H}{2\pi}$.   By Poincar\'e duality, for $M$ of dimension
4, 
$H^3(M,\Z)\cong H_1(M,\Z)$.   A general element of $H_1(M,\Z)$ can be defined by an embedded oriented circle $L\subset M$. 
In particular, $\y$ is ``Poincar\'{e} dual'' to such an $L$.   This means that $B$ is topologically trivial if restricted to the complement of $L$.
 Now we can perform a simple topological ``surgery,'' cutting
out of $M$ a small tubular neighborhood of $U$ of $L$ and then closing up the boundaries of the two resulting pieces.\footnote{Before making the surgery,
one can continuously vary $B$ so that it vanishes (rather than merely being topologically trivial) in $U\backslash L$ ($U$ with $L$ removed). Similarly one can continuously
vary $A$ to make it vanish in $U$.   This changes the action $J(A,B)$,
but the change is detrmined by the condition (\ref{ziml}). Once $A$ vanishes in $U$ and $B$ vanishes in $U\backslash L$, the surgery described in the text is straightforward.}
  This operation is a simple cobordism from $M$ to the disjoint union of
two manifolds.  One of those two manifolds, which we will call $M'$, is another copy of $M$, but with a topologically trivial $B$-field, and the second is  isomorphic to $L\times S^3$,
with a topologically nontrivial $B$-field, but with an $A$-field that automatically is topologically trivial (since $H^2(L\times S^3,\Z)=0$).    So the functional $J(A,B)$ on $M'$ or on $L\times S^3$
can be defined using (\ref{piml}), and the cobordism from $M$ to the disjoint union of $M'$ and $L\times S^3$ determines the difference $J_M-J_{M'}-J_{L\times S^3}$.   Putting these statements
together determines $J_M$.

 $A$-fields and $B$-fields both form additive groups, and the definition of $J(A,B)$ has made manifest that $J$ is linear in each variable:
\be\label{additivity}J(A_1+A_2,B)=J(A_1,B)+J(A_2,B),~~~~~ J(A,B_1+B_2)=J(A,B_1)+J(A,B_2). \ee

Having defined $J(A,B)$, the proof of the duality between the scalar and the two-form
requires knowing that the function $\exp(\i J(A,B))$ is a nondegenerate pairing between $A$ and $B$ in the following sense:
if $A$ is not gauge-equivalent to 0, then there exists $B$ such that $\exp(\i J(A,B))\not=1$.   In such a case, the additivity (\ref{additivity}) implies that the possible values 
of $\exp(\i J(A,B))$, for fixed $A$ but variable $B$,
form a subgroup of $U(1)$.  This subgroup is either all of $U(1)$ or else is a finite group $\Z_k$ for some $k>1$.   Either way, the average over $B$  of $\exp(\i J(A,B))$ is the same as the average over the subgroup and vanishes unless
$A$ is pure gauge.  That was the main step in deriving the duality between the massless scalar and the two-form gauge field.

The nondegeneracy of the pairing $\exp(\i J(A,B))$ is Poincar\'e duality of Cheeger-Simons differential characters.   It can be understood as a consequence of the following.
By virtue of the universal coefficients theorem of cohomology as well as  ordinary Poincar\'e duality, there is in general for any dimension $n$ and  any $k\leq n$ a perfect pairing
\be\label{relmo} H^k(M,\Z)\times H^{n-k}(M,U(1))\to H^n(M,U(1))=U(1). \ee
Perfectness of the pairing means that every homomorphism from either factor to $U(1)$ is obtained by evaluating this pairing with some element of the other factor;
in particular,
for every nonzero element of $H^k(M,\Z)$, this pairing is nontrivial for some element of $H^{n-k}(M,U(1))$, and vice-versa.  
Now let us try to prove that if $A$ is not pure gauge, then there exists $B$ such that $\exp(\i J(A,B))\not=1$.   We may as well assume that
$A$ is flat, since otherwise the formula $J(A,B)=\frac{1}{2\pi}\int_M F\wedge B$ makes manifest that $J(A,B)$ is nonzero for some topologically trivial $B$.
If $A$ is flat, then it corresponds to an element  $\zeta\in H^1(M,U(1))$.   Perfectness of the pairing (\ref{relmo}) with $k=3$ then provides an element $\y\in H^3(M,\Z)$
whose pairing with $\zeta$ is a non-identity element of $U(1)$.  A $B$-field whose characteristic class is equal to this $\y$ then has the desired property
that $\exp(\i J(A,B))\not=1$.

\section{Some More Details About the Path Integral}\label{zerograv}

Here we will describe a more explicit comparison between the path integral of the massless scalar $\phi$ and the two-form gauge field $B$ in four dimensions.
In fact, this has been treated in detail in the greater generality of a $p$-form gauge field and an $n-p-1$-form gauge field in $n$ dimensions \cite{DMW}.   Some aspects of what follows are explained more fully
in that paper.   The small fluctuations were previously analyzed in \cite{ST}.

The kinetic energy of the massless scalar field $\phi$ is the Laplacian $\Delta_0$ on zero-forms.  
The path integral over nonzero modes of $\phi$ is then roughly $1/\sqrt{\det'\Delta_0}$, where $\det'$ is the determinant with zero-modes removed.
But to be precise, as $\Delta_0$ is not dimensionless, its eigenvalues do not make sense as numbers and to make sense of its determinant requires
a choice of units.  
Actually, in quantization, to define the path integral measure precisely (see eqn. (\ref{nicemeasure})), one has to pick a parameter $\mu$ with dimensions of mass, effectively replacing
$\Delta_0$ with the dimensionless operator $\Delta'_0=\Delta_0/\mu^2$, whose  eigenvalues are numbers.
 Similarly, $\Delta_q$ will denote the Laplacian on $q$-forms and we define the dimensionless version $\Delta'_q=\Delta_q/\mu^2$.

The path integral for the non-zero modes of $\phi$ is a simple Gaussian integral, equal to $1/\sqrt{\det'\,\Delta'_0}$.
The path integral for small fluctuations of the two-form $B$ is similar, except that one has to include ghosts and antighosts in the quantization.   Indeed, in the case of a two-form gauge
field, the quantization can be carried out, as we explain presently, by adjoining to $B$ a pair of one-form fermionic fields of ghost number $\pm 1$ and three bosonic zero-form fields
of ghost numbers $-2,0,$ and 2.   The path integral for non-zero modes is thus
\be\label{pinoo}\left(\frac{1}{\det'\Delta'_2}\right)^{1/2} \left({\det}'\, \Delta'_1\right)  \left(\frac{1}{\det'\Delta'_0}\right)^{3/2}.  \ee

These regularized determinants obey two relations.   First, because of Poincar\'e duality, one has
\be\label{zinco}{\det}'\Delta'_q={\det}'\Delta'_{4-q},~~~~~~ q=0,\cdots,4.\ee
A second relation follows from the fact that the system of $q$-forms with $0\leq q\leq 4$ can be viewed as a supersymmetric system with supersymmetry generators $\d,\d^*$.
Every non-zero eiganvalue of $\Delta'_q$ is related by this supersymmetry to a non-zero eigenvalue of either $\Delta'_{q+1}$ or $\Delta'_{q-1}$ but not both.   This pairing of eigenvalues leads
to an identity
\be\label{windo}\frac{1}{\det'\Delta'_0} \,{\det}'\Delta'_1\,\frac{1}{\det'\Delta'_2} {\det}'\Delta'_3 \frac{1}{\det'\Delta'_4}=1.\ee 
Combining these two relations, one learns that 
\be\label{lindox}   \frac{1}{\sqrt{\det'\,\Delta'_0}}=\left(\frac{1}{\det'\Delta_2}\right)^{1/2} \left({\det}'\, \Delta'_1\right)  \left(\frac{1}{\det'\Delta'_0}\right)^{3/2}.  \ee
Thus the path integrals for non-zero modes of $\phi$ and $B$ are equal.

It remains to justify the claim about the fields that are introduced to quantize the two-form gauge field $B$.   This topic has been treated several times 
\cite{Townsend,Siegel,TM,ST}.
We will give a short explanation using the BRST approach to quantization.   In BRST quantization, the starting point is to introduce ghosts associated to the gauge symmetries.
For a two-form gauge field $B$, the ghosts are a fermionic 1-form $c_\mu$ of ghost number 1 with     BRST transformations $\delta B_{\mu\nu}=\partial_\mu c_\nu-\partial_\nu c_\mu$.   If  there is a redundancy among the gauge transformations, one further has to introduce ghosts for ghosts.   In the present case, there is a redundancy $c_\mu\to c_\mu+\partial_\mu\lambda$ for any $\lambda$.   So one adds
a bosonic field $\lambda$ of ghost number 2 with $\delta c_\mu=\partial_\mu\lambda$.    Thus up to this point, the BRST transformations are 
\be\label{inzo}\delta B_{\mu\nu}=\partial_\mu c_\nu-\partial_\nu c_\mu, ~~~\delta c_\mu=\partial_\mu\lambda,~~~ \delta \lambda=0.\ee
Clearly this is consistent with $\delta^2=0$.   Automatically the original classical action $I_{\mathrm{cl}}$ satisfies
\be\label{zillo} \delta I_{{\mathrm{cl}}}=0,\ee
since it depends only on $B$ and is gauge-invariant.

Beyond this point, one introduces trivial BRST multiplets, each  consisting of a pair of fields
 $\begin{pmatrix} \Lambda \cr K \end{pmatrix}$ with 
\be\label{winzo}\delta K=\Lambda,~~~\delta \Lambda = 0,\ee obviously still with $\delta^2=0$.
Here $K$ will have some ghost number $q$ and $\Lambda$ will have ghost number $q+1$.   Fields of even or odd ghost number are taken to be respectively bosonic or fermionic.
The upper component $\Lambda$ is often called a Nakanishi-Lautrup auxiliary field.\footnote{These  fields were originally introduced in covariant approaches to quantization
prior to the discovery of the BRST approach \cite{Nak,Lau}.   They were later reinterpreted as auxiliary fields in the BRST context.}  The lower component is often called an antighost.
Actually, quantization of the two-form gauge field is an example in which the traditional terminology concerning ``ghosts'' and ``antighosts'' is not very satisfactory, since as we will
see, one of the fields
that plays the role usually played by a ghost or antighost actually has ghost number zero.

After adding the trivial BRST mutliplets, one adds to the classical Lagrangian an additional term $\delta\Psi$, where $\Psi$ is a fermionic field of ghost number $-1$
that  may depend on the original classical fields,
the ghosts and ghosts for ghosts, and the trivial BRST multiplets, in an arbitrary fashion.   Because  $\delta^2=0$, $\delta\Psi$ is automatically BRST-invariant, and therefore the extended action
\be\label{brst}I_{\mathrm{BRST}} =I_{\mathrm{cl}} +\delta \Psi  \ee
is BRST-invariant.    $\Psi$ is called a gauge-fixing fermion.

The goal is now to pick the trivial BRST multiplets and the gauge-fixing fermion so that the extended action $I_{\mathrm{BRST}}$ has a nondegenerate kinetic energy for
all fields and therefore can be straightforwardly quantized.   Any such choice  leads to a satisfactory quantization, at least in perturbation theory.  The resulting quantum theory is invariant under changes in $\Lambda$, as long as $\Lambda$ is changed in such a way that the path integral remains always
well-defined \cite{Nielsen}.  (In general, a different choice of trivial multiplets or a modification of $\Lambda$ that cannot be reached by continuous variation of $\Lambda$
 without breakdown of the path integral might
lead to an inequivalent but equally valid quantization.)

For the two-form field $B$, this program can be carried out by adding three trivial BRST multiplets.  In describing them, we will use a superscript to indicate the ghost numbers, so the ghost fields
introduced previously will be denoted $c_\mu^{(1)}$ and $\lambda^{(2)}$.    The three trivial multiplets that we add are then a pair of one-forms $\begin{pmatrix} K_\mu^{(0)}\cr c_\mu^{(-1)}
\end{pmatrix}$ and two pairs of zero--forms $\begin{pmatrix} T^{(-1)}\cr \lambda^{(-2)}\end{pmatrix}$ and $\begin{pmatrix} T^{(1)}\cr \lambda^{(0)}\end{pmatrix}$.    The gauge-fixing fermion
is a sum $\Psi=\Psi_1+\Psi_2$.   The  first term is\footnote{For brevity, in the following we abbreviate $\frac{1}{h^2}\int\d^4x\sqrt g $ as just $\int$.} 
\be\label{firstone}\Psi_1=\int \left(-\frac{1}{2}c^{(-1)}_\mu K^{(0)\,\mu}+ \lambda^{(-2)} T^{(1)}\right).\ee  So
\be\label{zelfish} \delta\Psi_1=\int\left(-\frac{1}{2}K^{(0)}_\mu K^{(0)\,\mu}+  T^{(-1)}T^{(1)}   \right) .\ee
This is a nondegenerate quadratic function of the auxiliary fields $K^{(0)}$ and $T^{(\pm 1)}$.   Its role is to make it possible to integrate out the auxiliary fields.
The second part of the gauge-fixing fermion, which encodes the desired gauge condition, is 
\be\label{melfish}\Psi_2=\int\left(c_\nu^{(-1)}(D_\mu B^{\mu\nu}+\partial^\nu \lambda^{(0)}) -\lambda^{(-2)}D_\mu c^{(1)\,\mu}
\right).\ee
So
\begin{align}\label{ellish}\delta\Psi_2= &\int\left(K_\nu^{(0)}(D_\mu B^{\mu\nu}+\partial^\nu \lambda^{(0)})-c_\nu^{(-1)}\left(D_\mu(\partial^\mu c^{(1)\nu}-\partial^\nu c^{(1)\mu})
+\partial^\nu T^{(1)}\right)\right.\cr&\left.~~~~~ -T^{(-1)}\partial_\mu c^{(1)\mu} -\lambda^{(-2)}D_\mu\partial^\mu \lambda^{(2)}\right).
\end{align}
We will leave the reader to verify that in the combined action $I_{\mathrm{BRST}}=I_{\mathrm{cl}}+\delta\Psi$, after integrating out the auxiliary fields, every $q$-form field
of any $q$ and any ghost number has kinetic operator $\Delta_q$.  The field content consists of one two-form $B$, two fermionic one-forms $c^{(\pm 1)}$,
and three bosonic scalars $\lambda^{(-2)}$, $\lambda^{(0)}$, and $\lambda^{(2)}$,  as was assumed in arguing that  the Gaussian integral over small fluctuations is invariant
under duality.

It remains to discuss the integration over zero-modes.  The purpose of this discussion will be to learn how to interpret the factors  that appear in the Poisson resummation
formula (\ref{Poisson}) that relates the theta functions $\Theta_\phi$ and $\Theta_B$.    (This is treated in more detail in \cite{DMW}.)

 Zero-modes of the scalar field $\phi$ or of any of the fields $B,\psi^{(1)},\lambda^{(2)}$  give factors of  the coupling $f$ or $h$. We will explain this first for the scalar.
 Let us look more closely at the mode expansion of  $\phi$.   Let $\phi_n$ be the normalized eigenfunctions of the Laplacian $\Delta_0$, with
 eigenvalues $\gamma_n^2$.
 We make a mode expansion $\phi=\sum_n\alpha_n \phi_n$, with coefficients $\alpha_n$.   Then we perform a Gaussian integral over all coefficients $\alpha_n$ such that $\gamma_n\not=0$.
We claimed previously that the integral over non-zero modes of $\phi$ is $(\det'\,\Delta_0')^{-1/2}$, which is a regularized version of $\prod_n \frac{\mu}{\gamma_n}$.   To make
that true, the Gaussian integral over $\alpha_n$ should equal $\mu/\gamma_n$.
Since the kinetic energy for $\phi$ is actually $\frac{1}{2f^2}(\phi,\Delta_0\phi)=\frac{1}{2f^2}\sum_n \alpha_n^2\gamma_n^2$, to get the claimed result for
the Gaussian integral,  the measure in the integration over $\alpha_n$ has
to be
\be\label{nicemeasure}  \frac{\mu\,\d\alpha_n}{\sqrt{2\pi} f}. \ee
Here $\mu$ is an arbitrary mass, introduced to make this expression dimensionless.

Once we include a factor $\mu/\sqrt{2\pi}f$ in the measure for every non-zero mode, we have to also include that factor in the measure for every zero-mode as well.  The reason is that the measure
is supposed to be defined locally, and locally there is no way to distinguish zero-modes and non-zero modes.   One way to look at this is to imagine a lattice regularization in which the four-manifold
$M$ is approximated by $\beta_0$ points, where in the continuum limit $\beta_0\to\infty$.   To get the result (\ref{nicemeasure}) for the non-zero modes while defining the measure in a local fashion, 
we would include a factor $\mu/\sqrt{2\pi}f$ in the path
integral measure at each lattice point.   This will then give such a factor for the zero-modes as well.

For the non-zero modes, the factor of $1/f$ in the measure disappears upon doing the Gaussian integral, but for the zero-modes the factor $1/f$ in the measure actually leads directly to
such a factor  in the path integral.     We will write $b_q$ for the number of normalizable $q$-form zero-modes on the manifold $M$.   If $M$ is compact, the $b_q$ are the topological
Betti numbers of $M$.   If $M$ is not compact, they are an $L^2$ version of the Betti numbers.   For example, in the case of the wormhole spacetime, the topological Betti numbers
are $b_0=b_3=1$ with others vanishing, but the $L^2$ Betti numbers are $b_1=b_3=1$, with others vanishing.   In the particular case of a scalar field $\phi$, a zero-mode is a constant,
which is normalizable on a compact manifold but is not normalizable in a typical noncompact spacetime.  So $b_0=1$ if $M$ is compact, but typically $b_0=0$ if $M$ is not compact -- and in particular $b_0=0$  in the wormhole spacetime. When $\phi$ does have a constant zero-mode, that mode ranges over a circle,
since $\phi$ is an angular variable, so the integral over the zero-mode gives a finite result that includes a factor of $1/f$ from the measure.  Similarly,  because of Dirac quantization, zero-mode integrals for the $B$-field range over compact sets and give finite results.

Let us now include also the theta function $\Theta_f$ that comes from the sum over winding modes of $\phi$.   We found in eqn. (\ref{Poisson}) that
the relation between $\Theta_f $ and the analogous theta function  $\Theta_B$ for the $B$-field was $\Theta_f\sim f\Theta_B$.   This was for the case that the winding or flux modes
live in a rank 1 lattice; for the general case of rank $b_1$, we would have $\Theta_f\sim f^{b_1}\Theta_B$.  Including also the factors from the zero-mode measure,
 the scalar path integral is proportional to
\be\label{tellmot} f^{b_1-b_0}\Theta_B .\ee

As in \cite{DMW}, a straightforward way to determine the analog of this for the $B$-field is to consider precisely what it means to integrate over $B$-fields up to gauge equivalence.
We consider a lattice regularization with $\beta_0$ vertices, $\beta_1$ 1-simplices, and $\beta_2$ 2-simplices.   Thus the $B$-field has a total of $\beta_2$ modes, of which $b_2$ are zero-modes
and a certain number are gauge-equivalent to zero.   To count the modes that are gauge equivalent to zero, a first approximation is that there are $\beta_1$ generators of gauge transformations
on the lattice ($B\to B+\d\psi$), and $\beta_0$ generators of redundancies among these ($\psi\to \psi+\d\lambda$).   
But actually, of the $\beta_0$ potential generators of gauge redundancies, $b_0$ of them act trivially on $\psi$, and of the  $\beta_1-\beta_0+b_0$  modes of $\psi$ that
are not removed by these redundancies, $b_1$ act trivially on $B$.
 Thus the number of pure gauge modes of $B$ is actually $\beta_1-\beta_0 -b_1+b_0$, and the number of non-zero modes of $B$ mod pure gauge modes
  is $\beta_2-\beta_1+\beta_0-b_2+b_1-b_0$. 
 The integral over those non-zero modes will therefore give a factor of $h^{\beta_2-\beta_1+\beta_0-b_2+b_1-b_0}$.
 To define a local measure that gives a result independent of $\beta_0,\beta_1,$ and $\beta_2$ as those tend to infinity,
 we include a factor $h^{-1}$ in the measure for each two-simplex or vertex and a factor $h$
 for every one-simplex.   This cancels the factor $h^{\beta_2-\beta_1+\beta_0}$, and we are left
  over with a factor $h^{-b_2+b_1-b_0}=\left(\frac{2\pi}{f}\right)^{-b_2+b_1-b_0}$.   So the path integral of $B$ is equal to
 \be\label{nellmot} f^{b_2-b_1+b_0}\Theta_B \ee
 times factors independent of $f$.   
 
 The ratio of the $B$-field path integral to the scalar path integral is then $f^{b_2-2b_1+2b_0}=f^\chi$, where $\chi=b_0-b_1+b_2-b_3+b_4=2b_0-2b_1+b_2$ is the Euler
 characteristic of $M$, or an $L^2$ version of the Euler characteristic  if $M$ is not compact.  This result was first obtained in\footnote{The paper \cite{DMW} also contains a much more
 precise computation than we have explained here, fixing an overall constant that we have not determined (is the anomaly precisely $f^\chi$ or should $f$ be multiplied by a constant?).    
  In \cite{Witten}, electric-magnetic duality in four dimensions was derived in a similar fashion to what was presented here in section
 \ref{derivation}, and  the origin of an Euler characteristic anomaly was explained in  that framework.} \cite{Barbon,DMW}.  	If $M$ is compact, the Euler characteristic of $M$
 can be written as a curvature integral:
 \be\label{polo}\chi(M)=\frac{1}{32\pi^2}\int_M\d^4x \sqrt g\left(R_{ijkl}R^{ijkl}-4R_{ij}R^{ij}+R^2\right).\ee
 Therefore by adding to the action of the scalar theory
 a $c$-number term \be-\log f \frac{1}{32\pi^2}\int_M\d^4x \sqrt g\left(R_{ijkl}R^{ijkl}-4R_{ij}R^{ij}+R^2\right)\ee that depends only on the background metric,
 we can restore the equivalence between the scalar and $B$-field path integrals.     If $M$ is not compact, then the $L^2$ Euler characteristic of $M$ is given by the formula (\ref{polo}) plus a local boundary correction that is somewhat analogous to
 the Gibbons-Hawking-York boundary correction to the Einstein-Hilbert action.   That boundary term must be included in the action.
 
 Finally, consider rescaling the metric of $M$ by $g\to e^{2s}g$.   In the case of the wormhole spacetime, this rescales the wormhole radius $a_0$ by $a_0 \to e^s a_0$.
 The factor $1/a_0$ in the Poisson resummation formula (\ref{Poisson}) might appear to suggest that the relation between the scalar and two-form path integrals
 is affected by the rescaling of $g$.   That is not so, because, as  explained in \cite{DMW}, there are compensating factors from the rescaling of zero-mode measures,
 including the zero-modes of gauge generators and generators of gauge redundancies.

\vskip1cm
 \noindent {\it {Acknowledgements}}   I thank N. Arkani-Hamed, A. Heiderschee, V. Ivo, J.  Maldacena, and especially  G.  Moore for comments and discussions.
  Research supported in part by NSF Grant PHY-2514611.
 \bibliographystyle{unsrt}

\end{document}